\documentclass{vgtc}                          

\ifpdf
  \pdfoutput=1\relax                   
  \usepackage{graphicx}                
  \DeclareGraphicsExtensions{.pdf,.png,.jpg,.jpeg} 
\else
  \usepackage{graphicx}                
  \DeclareGraphicsExtensions{.eps}     
\fi%

\graphicspath{{figures/}{pictures/}{images/}{./}} 

\usepackage{microtype}                 
\PassOptionsToPackage{warn}{textcomp}  
\usepackage{textcomp}                  
\usepackage{mathptmx}                  
\usepackage{times}                     
\usepackage{cite}                      
\usepackage{tabu}                      
\usepackage{booktabs}                  
\usepackage{enumitem}

\onlineid{0}

\vgtccategory{Research}

\vgtcinsertpkg

\title{IEEE VIS Workshop on Visualization for Climate Action and Sustainability}

\author{
Benjamin Bach \thanks{e-mail: benjamin.bach@inria.fr}\\ %
\parbox{1.2in}{\scriptsize \centering Inria \& University of Edinburgh\\France/UK}
\and
Fanny Chevalier\thanks{e-mail: fanny@dgp.toronto.edu}\\ %
\parbox{1.5in}{\scriptsize \centering University of Toronto\\Canada}%
\and 
Helen-Nicole Kostis\thanks{e-mail: helen-nicole.kostis@nasa.gov}\\ %
     \parbox{1.5in}{\scriptsize \centering NASA/GSFC \& USRA/EfSI\\USA} %
\and 
Mark SubbaRao\thanks{e-mail: mark.u.subbarao@nasa.gov}\\ %
     \parbox{1.5in}{\scriptsize \centering NASA/GSFC\\USA} %
\and 
Yvonne Jansen\thanks{e-mail: YvonneJansenyvonne.jansen@cnrs.fr}\\ %
     \parbox{1.1in}{\scriptsize \centering CNRS\\France}
\and 
Robert Soden\thanks{e-mail: soden@cs.toronto.edu}\\ %
     \parbox{1.5in}{\scriptsize \centering University of Toronto\\Canada}
}

\abstract{
This first workshop on visualization for climate action and sustainability aims to explore and consolidate the role of data visualization in accelerating action towards addressing the current environmental crisis. Given the urgency and impact of the environmental crisis, we ask how our skills, research methods, and innovations can help by empowering people and organizations. We believe visualization holds an enormous power to aid understanding, decision making, communication, discussion, participation, education, and exploration of complex topics around climate action and sustainability. Hence, this workshop invites submissions and discussion around these topics with the goal of establishing a visible and actionable link between these fields and their respective stakeholders. The workshop solicits work-in-progress and research papers as well as pictorials and interactive demos from the whole range of visualization research (dashboards, interactive spaces, scientific visualization, storytelling, visual analytics, explainability etc.), within the context of environmentalism (climate science, sustainability, energy, circular economy, biodiversity, etc.) and across a range of scenarios from public awareness and understanding, visual analysis, expert decision making, science communication, personal decision making etc. After presentations of submissions, the workshop will feature dedicated discussion groups around data driven interactive experiences for the public, and tools for personal and professional decision making. 
} 

\begin{document}


\maketitle

\section{Motivation}

In mitigating climate change and providing for global sustainability, we are past the moment of just raising awareness: it is critical to adapt, achieve deep engagement, and empower everyone on earth in making decisions and taking actions~\cite{goldberg2020leveraging}. This includes a range of activities including researchers collecting and publishing empirical data around climate and sustainability; policy makers implementing regulatory laws to control carbon emissions and protect biodiversity; crisis responders intervening to aid communities in need and mitigate disasters (wildfires, landslisdes, water rationing, etc.); activist groups advocating for change; individuals considering lifestyle changes in terms of energy use, or product purchases to support sustainability; and businesses implementing the circular economy and making investment decisions. To support these processes, there is a vast range of data acquired, modeled, processed and archived by a multitude of government agencies, public, non-profit and private scientific organizations, that include deep historical records and on the ground observations, which require computing skills to be accessed, analyzed, interpreted, and acted upon.

\begin{figure}
    \centering
    \includegraphics[width=1\columnwidth]{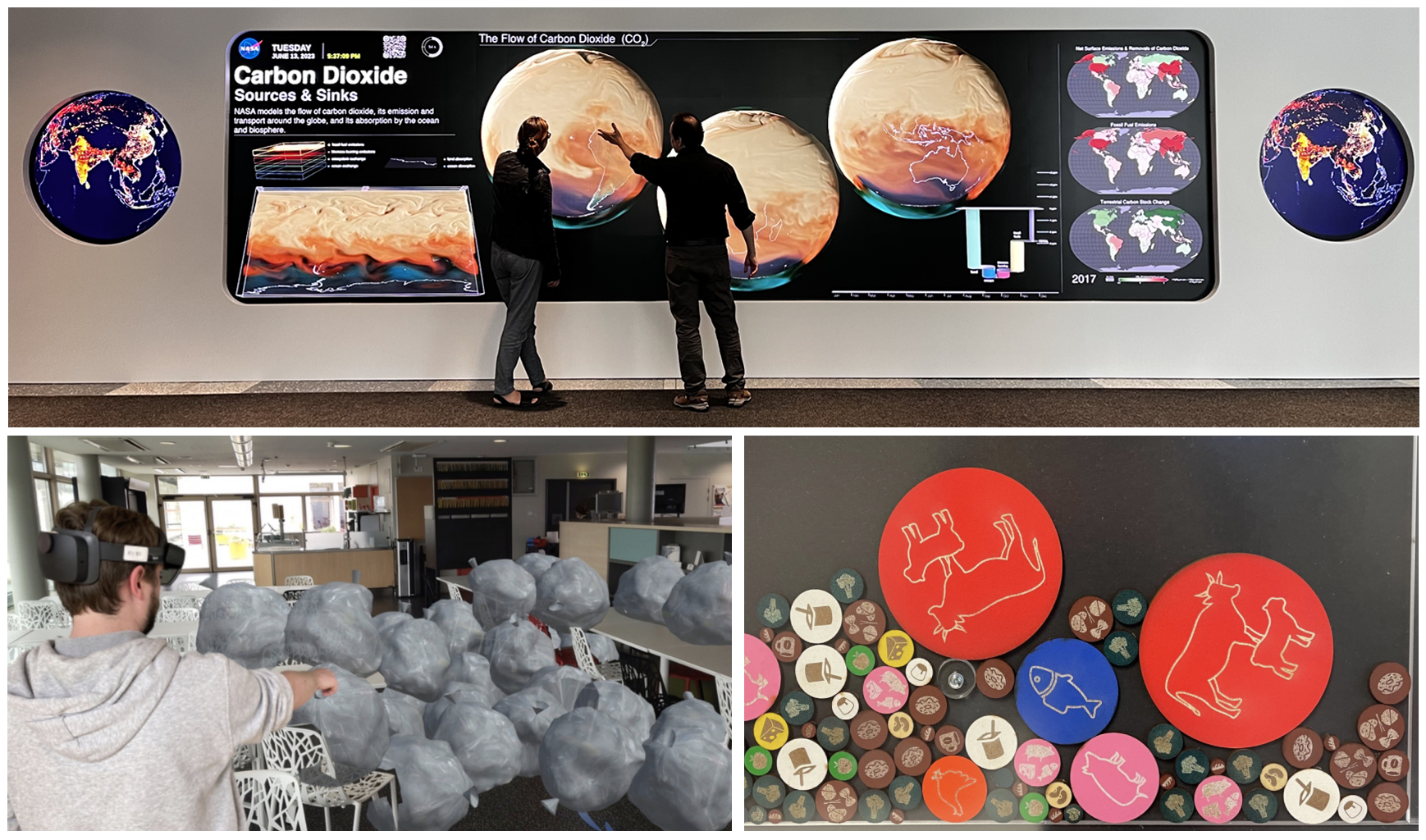}
    \caption{Top: Data dashboard at NASA's Earth Information Center 22ft LED Hyperwall public exhibit, located at NASA Headquarters in Washington, DC~\cite{kostis2022poster}; 
    Left: Augmented reality waste accumulation visualization ~\cite{assor2022augmented}; 
    Right: Participatory data physicalization of the impact of dietary choices~\cite{sauve2023edo}.}
    \label{fig:nasa-wall}
\end{figure}

 Data visualization is uniquely positioned to support many of these processes and has a long tradition in climate science. It also has produced a range of iconic images (e.g. climate stripes\footnote{{\small \url{https://showyourstripes.info/}}}, climate spiral\footnote{{\small \url{https://svs.gsfc.nasa.gov/5190/}}}, hockey-stick chart \cite{mann1999original} and heaps of public visualizations such as the 6th IPCC report~\cite{ipcc2022}, news\footnote{{\small \url{https://www.climatecentral.org}}}, art\footnote{{\small \url{https://art21.org/read/inigo-manglano-ovalle-climate}}}, and NASA’s Earth Information Center \footnote{{\small \url{https://earth.gov}}} dashboards and content providing easy access to first-hand data for the public (\autoref{fig:nasa-wall}-top)~\cite{kostis2022poster}.  To date, most of the work has been scattered across many different domains such as climate and environmental research, science communication~\cite{windhager2019inconvenient, windhager2019inconvenient-web}, domain specific applications, scientific storytelling~\cite{ma2011scientific}, and physicalization~\cite{sauve2023edo}. 
 Our community is yet to develop an overarching theory of practice and research agenda around climate visualization. To achieve this we need to examine  what exactly is specific about data visualization for climate action and sustainability and how working in this field can form and impact our research agendas. Climate and sustainability is such a huge---overwhelming, debated and arguably quickly evolving area---that we currently lack a clear understanding of why visualization for climate and sustainability justifies its own approaches and contributions, and what such are.

This workshop aims to reflect on how to link visualization research efforts around climate, sustainability, environmentalism~\cite{assor2022augmented}, circular economy, or food labeling~\cite{rondoni2021consumers,vlaeminck2014food} and foster a wider discussion about what visualization can contribute to accelerate action.

\section{Challenges and Opportunities}

The large scope and complex nature of the multi-faceted problem of visualization for climate action and sustainability is best addressed in an half-day dedicated, informal, interactive workshop that brings together diverse contributions, viewpoints, and reflections on the topic, followed by discussions.

The workshop encourages submissions around a broad variety of topics, challenges, and questions including (but not limited to): 

\vspace*{-1em}
\textit{
\begin{enumerate}[noitemsep]
\itemindent=-13pt
\item[] How to balance visual complexity, depth of information, and visual/data literacy?
\item[] How to understand abstract and widely unfamiliar scales of time, space, and numbers~\cite{chevalier2013using}?
\item[] How to encode, inform and present efficiently uncertain data? 
\item[] How to deal with heterogeneous data sets (spatial, temporal, relational, multidimensional, 3D, 4D, etc)?
\item[] How to make model projections more accessible and actionable for the general public?
\item[] How to tailor information and visualization to diverse audiences?~
\item[] How to work with policymakers and communities at risk?
\item[] How to create empathy with current and future people and populations~\cite{dragicevic2022information}?
\item[] How to support carbon accounting and monitoring systems?
\item[] How to track and monitor the circular economy?
\item[] How to support decision making on a personal as well as collaborative level?~\cite{mohanty2023save}?
\item[] How to foster engagement and participation on a community level?
\item[] How to leverage new and immersive technologies for analysis, communication, and awareness~\cite{dragicevic2022towards,besanccon2022exploring}?
\item[] How to train the creators of visualizations~\cite{bach2023challenges}?
\item[] How to prevent the misuse of visualization for this topic and provide for critical engagement?
\cite{bottinger2020challenges,bottinger2020reflections}?
\end{enumerate}
}

For some of these questions, we can start from the techniques and knowledge we have gained about visualization in general---others might require entirely new ways of thinking across information visualization, scientific visualization, analytics, illustration, information design, human-computer interaction, education, cognition, etc. Addressing these questions can lead to guidelines, collaborative platforms, visualization principles and techniques, toolkits, methodologies, visualization activities, analysis methods, interactive spaces and experiences, games, and potentially many more.

Research in climate action and sustainability can be a huge opportunity for (the) visualization (community) to unlock funding and make an impact, and at the same time enhance our knowledge about working with diverse audiences, across domain boundaries, build strong teams, solve real-world problems, and create collaborations with stakeholders of visualization that will inform our research for the coming decades. If not, we, as a community, are at risk of missing opportunities and losing agency within this highly dynamic space of prime importance. We argue that we need a structured effort to become aware and leverage the visualization super-powers~\cite{willett2021superpowers} such as our broad and yet deep knowledge and experience, our way of thinking, our enthusiasm for what we do and our networks, which this workshop aims to create momentum for.

\section{Previous related scientific initiatives}

This workshop directly builds on the {\textit{\textbf{Viz4Climate}}} \footnote{{\small \url{https://svs.gsfc.nasa.gov/Viz4Climate/}}} workshop from VIS 2022 and links into some other workshops. The Viz4Climate 2022 workshop had a strong focus on visual communication for climate (science), and two of its organizers participate in this proposal. The workshop featured a keynote by Ed Hawkins\footnote{{\small \url{https://edhawkins.org/}}}, a series of peer-reviewed 3min \textit{En}lightening presentations and a panel of academics, artists, and practitioners on the question \textit{What does High-Impact mean in the context of Visual Climate Science Communication?} This proposal goes further and wants to open up the topics beyond climate science communication and connects to a wider range of topics in visualization.

Related workshops held at the visualization conferences IEEE VIS and EuroVis include:

\begin{itemize}[noitemsep]
\item \textbf{Visualization for Social Good}, 2021-2023 includes many topics and submissions of relevance, but featured only few contributions around climate action and sustainability and none around a wider discussion of this topic.
\item \textbf{Visualization for Pandemic and Emergency Responses Workshop} (Vis4PandEmRes), 2023 discussed topics around how visualization researchers and visualization knowledge can support publich health emergencies, such as COVID-19. We aim to have such discussions around the climate emergency at our workshop.
\item \textbf{Discovery Jam}, 2017 featured an encounter between the visualization community and climate scientists and their particular problems. However, the idea was to sketch and discuss possibilities for collaboration rather than discussing the links between visualization and climate science more broadly. 
\item \textbf{EnvirVis} (Visualization in Environmental Sciences) Workshops hosted at EuroVIS yearly, with a strong focus on earth science and envrironmental applications.
\end{itemize}

Outside of the visualization conferences, we have
\begin{itemize}[noitemsep]
\item \textbf{Co-Creating Climate Futures} (2023) hosted at the MIT Media Lab and funded by NOAA (National Oceanic and Atmospheric Administration) explored the design and development of educational experiences around climate storytelling.\footnote{{\small \url{https://sites.google.com/media.mit.edu/grcworkshop2023/home}}}
\item \textbf{Making Communities More Resilient to Extreme Flooding Symposium} (2019), Earth from Space Institute (EfSI), with the aim of revolutionizing disaster response and community resilience through space technology. The 2-day symposium among others included sessions in visualizing flood risk and uncertainty and the role of data journalism.
\item \textbf{ACM CHI Workshop o Climate Change: Imagining Sustainable Futures}\footnote{{\small \url{https://sites.google.com/fbk.eu/hci-climate-change/home}}} which included topics around data physicalization, visualization, and sonification and has started discussing sustainability of their communities.\footnote{{\small \url{https://chi2023.acm.org/for-attendees/sustainability}}} \footnote{{\small \url{ https://chi2024.acm.org/2024/01/25/special-recognition-for-sustainable-practices}}}
\end{itemize}

\section{Workshop Goals}

The goals of this workshop are to 

\begin{itemize}[noitemsep]
\item \textbf{Collect research, examples, case studies, and experiences} around working with visualization in the broad area of climate and sustainability. 
\item \textbf{Discuss and explore the potential of how visualisation knowledge} and visualisation research can help address issues in climate and sustainability by supporting stakeholders in monitoring, decision making, communication, education and advocacy.
\item \textbf{Understand challenges and opportunities for general visualization research} (and practice), i.e., which are the problems the field needs to study and explore solutions for; and which are the gains we can drive for our general understanding of visualization from working on climate and sustainability. 
\item \textbf{Question the practices and implications} of our own work, e.g., with respect to the use of technology, their environmental impact (e.g., computing power, physical materials), or conferences and travel~\cite{lee2023only}. 
\item \textbf{Build a community of practitioners and researchers} across all fields of visualization and associated areas (machine learning, climate science, education, advocacy, ...). We particularly encourage (future) collaboration among the attendees.
\end{itemize}

\section{Submission Formats}

\begin{itemize}[noitemsep]
\item \textbf{Regular publications (up to 8 pages)}, including surveys, case study reports, and original research that advance the field.
\item \textbf{Short position statements / semi-public notes (2-3 pages)}, featuring opinions, experiences, lessons learned, and reflections, which together will contribute diverse viewpoints that enhance collective understanding of, and encourage discussions around the theme of the workshop. 
\item \textbf{Pictorials+interactive demos (1 page + supplemental materials)} about topics in climate action and sustainability. These can come in the form of infographics, data-driven animations, data comics, data physicalizations, installations, and VR/AR/XR experiences etc. Submissions to this category encompass a 1-page write-up together with the piece itself where possible (e.g. URL to the demo or short video clip) or materials featuring the details (e.g. videos or photographs of a data physicalization or installation, space and technical requirements, etc...). 
\end{itemize}

Each of the submissions will be peer-reviewed by at least two members of the IPC as well as one of the workshop organizers. All accepted submissions will be made available to workshop attendees and IEEE Vis participants.

\section{Activities and Tentative Schedule}

Our workshop is planned as a half-day hybrid workshop.

\textbf{Session 1 (before break, 1h15)}---Welcome and Scoping (10min), followed by lightning  presentations (3mins each) of accepted submissions/installations. Questions  and answers will follow presentations. 

\textbf{Session 2 (after break, 1h15)}---Brief introductions to get in person and remote  participants to get acquainted with other attendees. Form discussion groups around topics proposed by the organizers and the audience. Each discussion group is moderated by one of the organizers. Tentative topics include:

\begin{itemize}[noitemsep]
    \item How to build immersive spaces and experiences for climate storytelling?
    \item How to raise climate education, awareness and change behavior on a personal level?
    \item How to support sustainability monitoring, community engagement, and  participatory decision making?
 \end{itemize}

The above questions are examples of possible directions. The organizing committee will refine the list of topics for discussion using the submissions as indicator of participants' interest, and input from attendees. 
Each \textit{working group} will be tasked to compile a small (visual) report of their working group discussion including opinions, examples, challenges etc. to share during the workshop.

\section{Workshop Support Requirements}

The workshop proposed will be a hybrid one. It will be available on-site at the conference center and will be also available for virtual attendees and presenters. We will need the following support: 
\begin{itemize}[noitemsep]
\item Room and AV support at the conference (on-site). 
\item Videoconferencing options and support for live streaming to, and participation from remote attendees. 
\item Dedicated Discord channel for the workshop.
\item Access to PCS for submissions and the review process.
\item 2-4 poster slots.
\item Inclusion of the workshop materials or proceedings in the downloadable conference proceedings for dissemination to attendees of the conference and in IEEE Xplore.
\item IEEE Xplore Digital Library publication of the workshop proceedings.
\item Food and beverages for on-site attendees (one coffee break).
\end{itemize}

We are committed to promote policies and practices of equity, diversity and inclusion involved in the process of selecting submissions to be presented at the workshop, as well as that affecting participation. As such, we will work with the workshop chairs to support underprivileged participants who may face obstacles which would prevent their attendance to the workshop for cases which go beyond the default available funds for special purposes .

\section{Pre-Workshop Organization Timeline}

Pending acceptance, the organization tentative timeline for the workshop is as follows:
\begin{itemize}[noitemsep]
\item \textbf{March 20, 2024:} Call for Participation goes live. 
\item \textbf{June 1, 2024}: Submissions deadline. 
\item \textbf{June 30, 2024}: Review deadline.
\item \textbf{July 10, 2024:} Author notifications.
\item {\textbf{July 25, 2024:}} Submission camera ready deadline.

\end{itemize}

Workshop organizers aim to advertise the call for submissions and the final program through social media and mailing lists among colleagues from various communitie including: IEEE VIS, ACM CHI, ACM SIGGRAPH, ACM DIS, PacificVis, Eurovis, Digital Humanities, Art+Design, Tableau, etc.

\section{Intended Outcomes}
The workshop's primary outcomes are:
\begin{itemize}[noitemsep]
\item a collection of contributions around visualization for climate action and sustainability centralized on the workshop website;
\item community building through novel connections, sharing of work-in-progress, research interests, and plans;
\item collaborations inspired by workshop presentations and community building discussions;
\item actionable insights which will inform defining challenges and opportunities, guidelines, and best practices on the topic.
\end{itemize}

Indicators of success of this workshop include the number of submissions and attendees, which we will report to the workshop chairs; the diversity and quality of accepted submissions which one can subjectively assess by looking at the contributions made available on the workshop website; as well as events, papers and follow-up discussions that come out of this workshop that attendees make us aware of.

\section{Organizing Committee}

\textbf{Benjamin Bach} ({\small \url{http://benjbach.net}}) is a researcher at Inria (France) and an Associate Professor in Visualization at the University of Edinburgh (UK). His research creates and evaluates techniques and methods to empower people using visualizations for their private and public agendas. Benjamin has been working on network visualization, visualization in immersive environments, data-driven storytelling, and education for visualization literacy. He has been organizing several workshops at the IEEE VIS (VisActivities, VisGuides, EduVis). 

\textbf{Fanny Chevalier} ({\small \url{http://fannychevalier.net/}}) is an Assistant Professor in Computer Science and Statistics at the University of Toronto, Canada and Knight of the France's Order of the Academic Palms. Her research is interested in methods and tools supporting visual analytics and creative activities, with primary focus on interactive visualization for the visual exploration of rich and complex data, visualization education and statistical communication, and computing tools supporting the flow of creativity. Fanny has been working on applications spanning healthcare, social network analysis, education, cinematography, and digital arts.

\textbf{Helen-Nicole (Eleni) Kostis} ({\small \url{https://science.gsfc.nasa.gov/sed/bio/helen-nicole.kostis-1})} leads data dashboards and visualizations for NASA's \href{https://earth.gov/}{Earth Information Center} hyperwall displays. Based at the \href{https://svs.gsfc.nasa.gov/}{Scientific Visualization Studio (SVS)} Eleni develops, designs and produces data-driven media, conduits, and experiences with the goal to communicate complex climate phenomena and research findings to the scientific communities and the public. In 2010, with fellow team members Eleni brought to life the NASA Visualization Explorer---a first of its kind conduit that released two visualization stories per week. During her graduate studies she developed and designed tele-immersive environments using high-speed experimental networks. Eleni is the recipient of NASA’s Exceptional Achievement Award for Outreach and co-author of \href{https://doi.org/10.1007/978-3-030-34444-3}{the Foundations of Data Visualization.}

 \textbf{Mark SubbaRao} ({\small \url{https://marksubbarao.github.io/}}) leads the \href{https://svs.gsfc.nasa.gov/}{NASA’s Scientific Visualization Studio}, a group tasked with visualizing NASA science results for public audiences. Before joining NASA, Mark spent 18 years at the Adler Planetarium in Chicago, where he produced planetarium shows and designed museum exhibits featuring data-driven scientific visualizations. During 2019-2020 Mark served as President of the International Planetarium Society (IPS), where he spearheaded the ‘\href{https://www.ips-planetarium.org/general/custom.asp?page=data2dome}{Data to Dome’’} initiative - an effort to prepare the planetarium community for the big data era. Before that he worked at the University of Chicago where he was part of a team that created the largest 3D map of the Universe, the Sloan Digital Sky Survey. Mark's visualizations have been widely featured in print, TV, museums, and even projected on the sides of buildings.

\textbf{Yvonne Jansen} is a researcher in human-computer interaction and data visualization at the French Center for Scientific Research (CNRS). Her research interests include data physicalization, situated and embedded visualization and more generally how to help people make sense of complex data and phenomena, especially in the context of climate change mitigation. She previously organized multiple workshops at different conferences, including at IEEE VIS and ACM CHI.

\textbf{Robert Soden} ({\small \url{http://robertsoden.io/}}) is an Assistant Professor at the University of Toronto working on climate informatics, human-centered computing, and science and technology studies. His research uses a range of ethnographic, participatory, and design research methods to evaluate and improve the technologies we use to understand and respond to environmental challenges like disasters and climate change. Robert has been organizing several workshops at the ACM CHI conference, including the \textit{HCI for Climate Change: Imagining Sustainable Futures} in 2023.

\bibliographystyle{abbrv-doi}

\bibliography{main}
\end{document}